# Skyrmion flow in periodically modulated channels

Klaus Raab[1,‡], Maurice Schmitt[1,‡], Maarten A. Brems[1], Jan Rothörl[1], Fabian Kammerbauer[1], Sachin Krishnia[1], Mathias Kläui[1, a)], Peter Virnau[1, b)]

[1] Institut für Physik, Johannes Gutenberg-Universität Mainz, Staudingerweg 7, 55128 Mainz, Germany

‡ These authors contributed equally to the study.

**E-mail addresses (Corresponding authors)** a) klaeui@uni-mainz.de, b) virnau@uni-mainz.de

**Abstract**

Magnetic skyrmions, topologically stabilized chiral magnetic textures with particle-like properties have so far primarily been studied statically. Here, we experimentally investigate the dynamics of skyrmion ensembles in metallic thin film conduits where they behave as quasi-particle fluids. By exploiting our access to the full trajectories of all fluid particles by means of time-resolved magneto-optical Kerr microscopy, we demonstrate that boundary conditions of skyrmion fluids can be tuned by modulation of the channel geometry. We observe as a function of channel width deviations from classical flow profiles even into the no- or partial-slip regime. Unlike conventional colloids, the skyrmion Hall effect can also introduce transversal flow-asymmetries and even local motion of single skyrmions against the driving force which we explore with particle-based simulations, demonstrating the unique properties of skyrmion liquid flow that uniquely deviates from previously known behavior of other quasi-particles.

*Introduction -* Skyrmions are chiral magnetic quasiparticles[1–6] that have attracted significant attention since their discovery[6] in the fields of magnetism and spintronics owing to their topologically enhanced stability[7–9]. Once nucleated, these non-trivial spin textures can be manipulated and moved by using low-power spin-orbit torques induced by electric current[10–13]. These properties make them interesting for potential application in spintronics[1,4,14–25] but also as a model system in statistical physics, as they can be considered a highly tunable, virtually ideal two-dimensional system[26]. Dynamic features like size-control[15,27], stochastic Brownian motion[15,28–30] and skyrmion-skyrmion-[31–33], as well as skyrmion-boundary[31–35] repulsion in patterned structures grant them high experimental value to model rheological 2D phenomena like lattice ordering[6,34], phase transitions[34,36,37] and fluid motion[38,39]. In this context, skyrmion flow is the semi-synchronous, driven motion of individual skyrmions in a dense state[39], confined by other skyrmions into a low density lattice.

Classical incompressible Newtonian fluids in laminar flow obeying the Navier-Stokes equation exhibit Poiseuille flow[40] for which the velocity flow profile along the cross-section of a simple confining geometry with flat walls is parabolic. In the context of Newtonian fluids in laminar flow, the frictional interaction between fluid particles and the wall of the channel leads to the observation of the lowest velocity at the channel's walls, with the highest particle velocity occurring in the middle of the channel. The fulfilment of the no-slip boundary conditions occurs as the velocity of fluid particles at the channel's wall approaches zero. In cases where the velocity decreases toward the boundaries but does not reach zero, the system satisfies partial-slip conditions[41].

The flow of magnetic skyrmions induced by spin-orbit torques exhibits fluid-like characteristics analogous to, e.g., colloidal fluids[42]. Similar to the latter, skyrmions also move in a rough energy landscape originating from defects in magnetic thin films, so-called pinning sites[43]. In a hypothetical perfect, pinning-free skyrmion sample with straight walls and without stray fields, no interactions exist that slow down skyrmions near the boundary of the system and thus leads to slip boundary conditions. No-slip boundary conditions can be fulfilled artificially in experiments if channels with rough confining walls are used, e.g., by triangular modulation of edges. Notably, this means that the boundary conditions of skyrmion fluids can be controlled by the artificial roughness of the structure edge. Comparable rough geometries have been used to tune the flow in simulations of colloidal systems[44,45]. As the constituent particles of a skyrmion fluid can also be tracked individually in space and time, this makes skyrmion fluids compelling systems to study flow phenomena in two dimensions. However, in contrast to colloids, so far little has been studied, in particular experimentally, about the flow properties of skyrmions.

In this study, we investigate rheological characteristics and dynamics of skyrmion systems and their viability as a dynamical 2D model system. We determine the influence of boundary conditions on skyrmion channel flow experimentally and observe drastically varying velocity profiles. With quasi-particle simulations we explore conditions beyond the current capability of our experimental setup and demonstrate asymmetric channel flow inaccessible to conventional colloidal systems showing how additional complexities of skyrmion dynamics result from the skyrmions' topological spin structure.

*Experimental skyrmion velocity flow profiles* – We employ experimental measurements on a thin film sample deposited by magnetron sputtering with the following layers beginning with the seed layer on a thermally oxidized Si substrate: Ta(5)/Co$_{20}$Fe$_{60}$B$_{20}$(0.95)/Ta(0.09)/MgO(2)/Ta(5) (respective thicknesses in nanometers in parentheses and relative stoichiometric concentration in percentage). More information on manufacturing of the sample is given in Appendix A. We record the skyrmion dynamics using magneto-optical Kerr effect (MOKE)[46] microscopy with differential imaging technique at a frame rate of 16 frames per second. More information on the experimental protocol is given in Appendix B. Spin-orbit torques induced by electrical currents at the Ta/CoFeB interface cause controllable skyrmion motion[12,47] along the current flow with skyrmion velocities in the µm/s regime at low electric current densities of the order of $10^7 \frac{A}{m^2}$ [16]. The samples' skyrmion-edge and skyrmion-skyrmion repulsion confine the particles in the available space, causing a relatively dense skyrmion state when nucleated[15,33,34] (see Fig. 1. a)). To examine the impact of the imposed boundary conditions on the fluid-like behavior of skyrmions, we have fabricated channels with varying confinements. The confinements include straight and thus sharp (Fig. 1 (a)) wire edges, and diametrically zigzag modulated and thus rough edges (Fig. 1 (c, e)).

To understand the type of flow, we first determine the velocity flow profile of skyrmions from trajectories based on frame-to-frame displacements. Fig. 1 a) shows one frame with tracked skyrmions encircled in red, while Fig. 1 b) displays the resulting experimental velocity flow profile for an applied current density of $5.7 \cdot 10^7 \frac{A}{m^2}$ along the width of the wire. In a straight wire, the velocities exhibit a flow that is not very significantly influenced by the edge of the geometrical confinement. Hence, this setup corresponds to slip conditions. Discrepancies observed between experimental flow and a completely homogeneous skyrmion velocity profile likely arise from pinning effects as discussed in Supplementary Figure 1. Conversely, the skyrmion velocity profile in the modulated channel with an inner width of 80 µm (see Fig. 1 d)) clearly shows the influence of the modulation: While velocities in the inner part are almost constant, velocities in the outer parts within the triangular modulation rapidly decrease towards the edge, as the skyrmion motion is strongly hindered due to edge repulsion and lower current density. Thus, we can effectively impose partial-slip condition with these boundary modulations.

When we decrease the width of the channel but keep the edge modulation, the flattened region in the center of the velocity profiles can be minimized, or even avoided, as seen in Fig. 1 e and f). In this experiment, the distance between opposing triangle edges is halved to a 40 µm channel width, while the current density stays the same, resulting in an almost parabolic velocity profile, resembling that of a classical fluid. Note, however, that this is a finite-size effect and only occurs in narrow channels while the repulsive interactions stay unaltered.

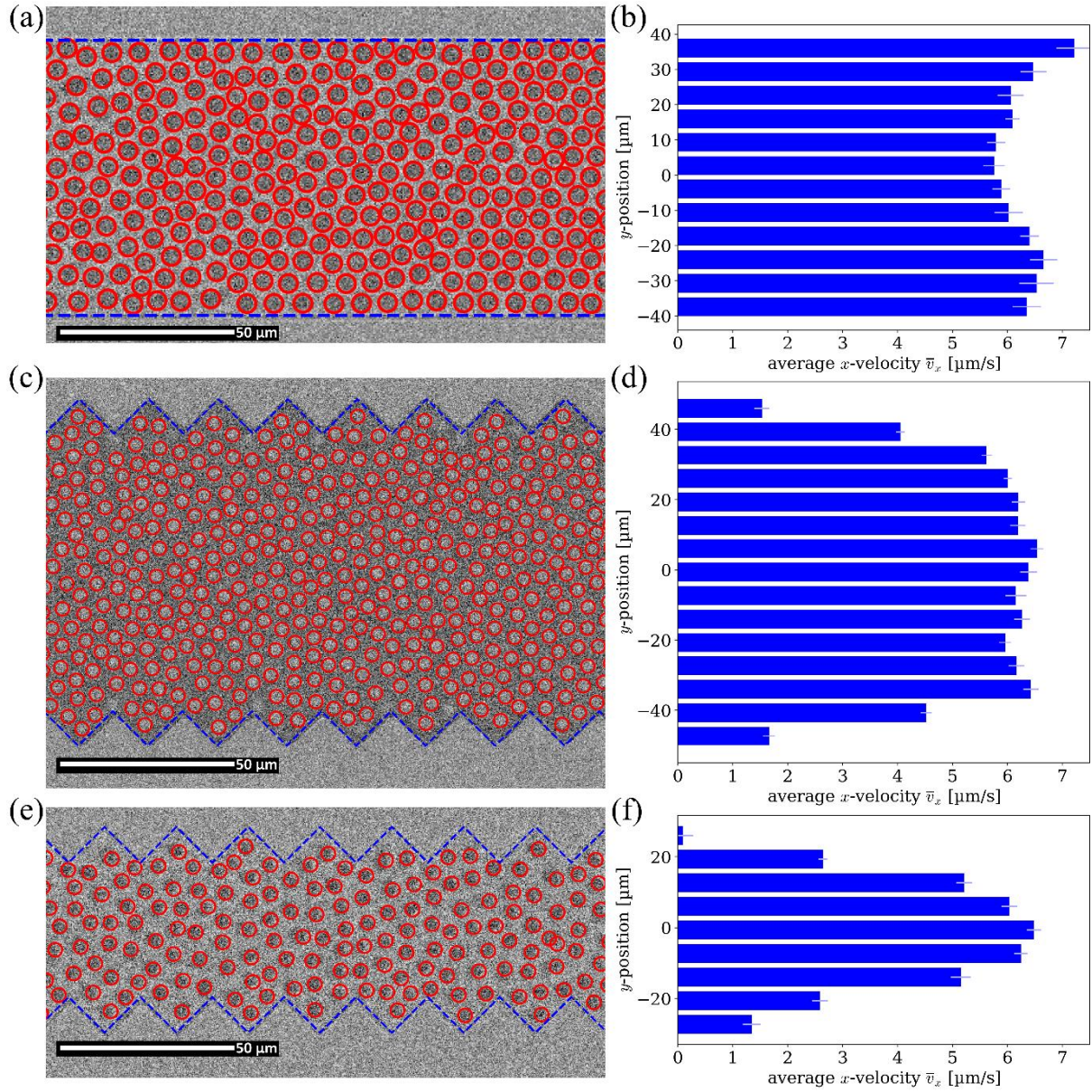

**Figure 1:** a), c) and e) Single experimental Kerr microscopy frame with skyrmions encircled in red in channels with straight (a) and modulated edges (c, e). Skyrmions are tracked frame by frame and their trajectories are used to extract their respective velocity profiles in x-direction (along the wire). The blue dashed lines help illustrate the channels edge/confinement, which is not visible due to difference imaging. b), d), f) Corresponding velocity profiles; showing the average velocity in µm/s of tracked skyrmions across the channel width (y-axis). Lines at the tip of each histogram bar represent error bars.

To rationalize and better understand the observed behavior, we perform particle-based simulations parametrized to reproduce conditions leading to Fig.1c) and d). Experimental data of rectangular current pulses leads to effective skyrmion Hall angles of $\theta_{skHA} = (0.58 \pm 0.13)°$ [48] and hence corresponding simulations can safely assume $\theta_{skHA} = 0$. In a second step, we explore scenarios beyond the capabilities of our current experimental setup and investigate skyrmions with spin structures that go beyond our experimentally realized spin structures leading to strong flow asymmetries arising from non-zero skyrmion Hall angles.

***Thiele Model Simulations*** – Quasi-particle simulations are based on the common form of the Thiele framework[49–52] with isotropic dissipation, which is used to describe skyrmions as quasi-particles:[53,54]

$$-\gamma\vec{v} + G^z \times \vec{v} + \vec{F} = 0$$

with friction coefficient $\gamma$, instantaneous velocity $\vec{v}$ induced by the total force $\vec{F}$, and gyrotropic vector $G^z = (0,0,G)$ responsible for the skyrmion Hall effect[a]. The corresponding term is set to zero for our first set of simulations while our study of flow asymmetries requires a finite Magnus force. Contributions to the total force $\vec{F}$ are a driving force $\vec{F}_D$, replicating spin-orbit torque induced motion through a confining geometry, a random force $\vec{F}_R$ to mimic Brownian diffusive motion observed in skyrmions, as well as purely repulsive skyrmion-skyrmion and skyrmion-boundary interactions based on previous results of iterative Boltzmann inversion[33]. Note that conventional colloidal systems are also simulated with a similar overdamped Langevin equation which, however, typically does not contain a gyro tropic term required to induce flow asymmetries[55,56] and thus skyrmion flow dynamics exhibits additional complexity not found in conventional colloids.

Further details on the simulation protocol are given in Appendix C. As there is no inertia in the model, skyrmions in the middle of the confining geometry approximately flow with velocity $|\vec{v}_{max}|$ directly induced by the driving force $\vec{F}_D$ in the center:

$$|\vec{v}| = |\vec{v}_{max}| = \frac{|\vec{F}_D|}{\gamma}.$$

The applied driving forces directly follow current density distributions corresponding to each experimental sample, which are obtained using the COMSOL software package[57]. One exemplary current distribution of a sample with modulated edges is given in Supplementary Figure 2.

***Simulations of symmetric channel flow*** – First, simulations with a negligible skyrmion Hall effect ($\gamma$ = 1, G = 0) were performed and reproduce the creep regime behavior observed in Fig. 1c), d). The average velocity only decreases in the immediate vicinity of the modulation, while skyrmions in the inner part of the channel flow almost homogeneously as they are accelerated back to $|\vec{v}_{max}|$ immediately after collisions due to lack of inertia. The resulting flat profile in the center of the channel (Fig. 2 red) is in good agreement with corresponding experimental results (Fig. 2 blue) confirming the viability of our approach. In comparison, skyrmion flow in a straight channel without modulation (not shown here) is not significantly hindered and the velocity profile is flat even close to the confining boundaries. We conclude that the flow behavior does not resemble that of a classical fluid but is indeed very similar to behavior expected for other quasi-particle systems such as colloids as both can be modelled using similar equations of motion.

---

[a] It is important to ensure that the dissipative force opposes the total force, whereas the sign of the Magnus force does not affect the results except for a spatial reflection. In some of our previous publications[32,33,35] there have been typos concerning the sign of the dissipative force. All simulations have however been using the correct equations.

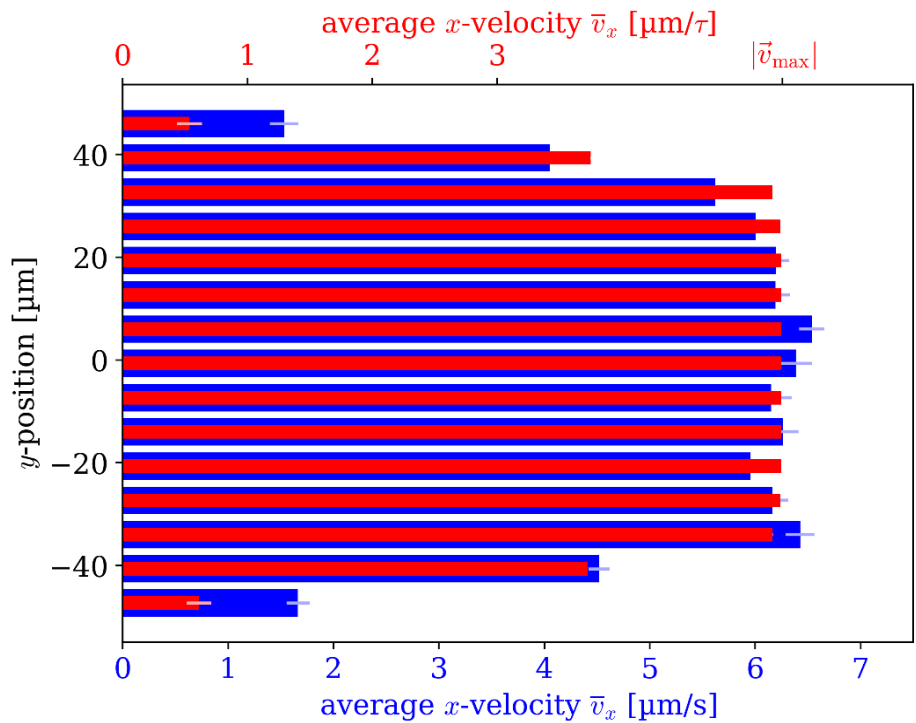

**Figure 2:** Skyrmion flow velocity profiles of the modulated channel of 80 µm width. Comparison of experimental (blue) and simulated (red) velocity profiles. Both profiles become flat in the inner part of the channel.

*Simulations of asymmetric channel flow and local backflow –* While so far, our simulations did not consider the Magnus term, we next include this term to account for the skyrmion Hall effect. With non-zero skyrmion Hall angles, the mirror symmetry of the velocity flow profiles with respect to the center of the channel is broken and driving forces push skyrmions preferentially to one side of the sample. This does not only lead to asymmetric skyrmion flow, but reinforced interactions with modulated edges enable interesting and more complex flow phenomena inaccessible to conventional colloidal systems. One of these examples we discuss below.

In a particular confining geometry with triangular modulation, the average velocity of skyrmions flowing near the bottom edge of the geometry becomes negative in simulations with a skyrmion Hall angle of $\theta_{skHA} = 15°$, i.e., skyrmions propagate opposite to the applied force. The striking asymmetry in the simulated velocity flow profile between the upper half of the confining geometry and the lower half is shown in Fig. 3a). Notably, the velocity increases significantly towards the upper edge, while the velocity towards the lower edge decreases and even becomes negative. These asymmetries are purely caused by the skyrmion Hall effect, since all forces and the confinement are symmetric in our simulations and there are no pinning-effects in this system. Similar asymmetries might be observable in experiments with larger applied current densities where the skyrmion Hall effect is not suppressed[48] and the effects of pinning are overcome. However, tracking the dynamics of skyrmions within the viscous flow regime (velocities on the order of m/s)[48] while a direct current is applied is currently not possible in our experimental setup. The negative velocity at the bottom edge of the confining geometry in these simulations is caused by a dynamic effect in which one skyrmion forces another one into the modulation space, which then moves against the flow direction along the contour of the modulation. As sketched in Fig. 3b), a skyrmion (marked in red) can be shoved into the space between two modulating triangles by a skyrmion located directly behind it (marked in blue). Notably, this only occurs frequently at the bottom edge of the confining geometry since skyrmions are constantly pushed upwards by the skyrmion Hall effect, thus leaving more room for a skyrmion to be pushed between two triangles at the lower edge of the confining geometry. After following the contour of the modulation, the skyrmion marked as red realigns with the flowing skyrmions in the bulk effectively switching positions with the skyrmion marked as blue. During this process, the red skyrmion is pushed against the applied current induced spin-orbit torque, which is lower inside the triangular modulation (see Suppl. Fig. 2), thus resulting in a negative velocity for that particular bin. Note, however, that this effect does not lead to an effective backflow as skyrmions are propelled in the flow direction as soon as they move out of the confining modulation.

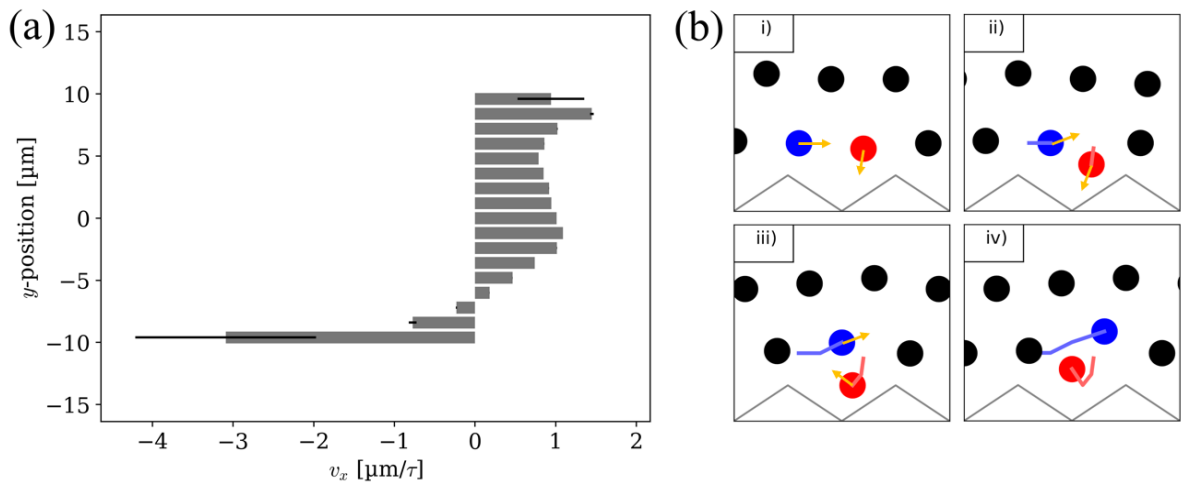

**Figure 3:** a) Simulated velocity flow profile of a 20 μm channel with oblate modulation (see b)) and a skyrmion Hall angle set to $\theta_{skHA} = 15°$, resulting in an asymmetric velocity flow profile. Notably, the average skyrmion velocity at the lower edge of the confining geometry is negative, implying that skyrmions in this region flow opposite to the applied driving force. b) Sketch of a dynamic process

which occurs at the lower edge of the confining geometry, leading to negative velocities on average. Skyrmions experiencing a force from left to right will also be pushed upwards due to the skyrmion Hall effect. As a result, a skyrmion (marked red) can be pushed into the space created between the flowing skyrmions and two triangle tips (i-ii). (iii-iv) Then the red skyrmion slightly moves against the applied current and finally moves back in line with the flowing skyrmions, behind the skyrmion it was initially ahead of.

*Conclusion* **-** In this paper, we determine the characteristics of skyrmion quasi-particle flow in a two-dimensional channel by combining experiments and simulations. We demonstrate that a large variety of flow profiles can be obtained by tuning the confining boundaries. A regular triangular modulation hinders movement of skyrmions along the boundaries effectively imposing partial-slip conditions, while flat boundaries provide slip conditions. In the modulated case we obtain an almost parabolic flow profile for small channels and driving forces, while velocities flatten in the middle of wider channels. On a fundamental level, skyrmion quasi-particles can thus share many transport characteristics with other macroscopic particles in overdamped scenarios like conventional colloids. Unlike the conventional quasi-particles, however, asymmetries arising from Magnus forces enable a whole new variety of phenomena including local backflow which we demonstrate with particle-based simulations. This together with the ability to adjust sizes and densities of skyrmions on the fly underscores their potential as a highly tunable model system for studying fundamental aspects of forces and transport in statistical physics.

**Code availability**

The computer codes used for data analysis are available upon reasonable request from the corresponding authors.

**Data availability**

The data supporting the findings of this work are available from the corresponding authors upon reasonable request.

**Acknowledgements**
P. V. acknowledges helpful discussions with M. Lukacova. The work was supported by the Deutsche Forschungsgemeinschaft (DFG, German Research Foundation) projects 403502522 (SPP 2137 Skyrmionics), 49741853, and 268565370 (SFB TRR173 projects A01 and B02), TopDyn and ERC-2019-SyG no. 856538 (3D MAGiC). P. V., M. S., M. B. and J. R. acknowledge funding from DFG

SFB TRR 146 (project #233630050). M. B. is supported by a doctoral scholarship of the Studienstiftung des deutschen Volkes.

**Author Contributions**

M. K. and P. V. devised and supervised the study. F. K. sputtered the sample. K. R. designed and fabricated the sample and carried out the measurements. K. R. and M. S. tracked and analyzed recorded videos. M. S. performed the simulation and analyzed the flow data with the help of M. B. and J. R.. The manuscript was prepared by M. S., K. R., M. B. and P. V.; all authors commented on the manuscript.

**Competing Interests**

The authors declare no conflict of interest.

**References**

1. Fert, A., Reyren, N. & Cros, V. Magnetic skyrmions: advances in physics and potential applications. *Nat. Rev. Mater.* **2**, 17031 (2017).

2. Bogdanov, A. N. & Panagopoulos, C. Physical foundations and basic properties of magnetic skyrmions. *Nat. Rev. Phys.* **2**, 492–498 (2020).

3. Jiang, W. *et al.* Skyrmions in magnetic multilayers. *Phys. Rep.* **704**, 1–49 (2017).

4. Finocchio, G., Büttner, F., Tomasello, R., Carpentieri, M. & Kläui, M. Magnetic skyrmions: from fundamental to applications. *J. Phys. Appl. Phys.* **49**, 423001 (2016).

5. Wiesendanger, R. Nanoscale magnetic skyrmions in metallic films and multilayers: a new twist for spintronics. *Nat. Rev. Mater.* **1**, 16044 (2016).

6. Mühlbauer, S. *et al.* Skyrmion Lattice in a Chiral Magnet. *Science* **323**, 915–919 (2009).

7. Pokrovsky, V. L. Properties of ordered, continuously degenerate systems. *Adv. Phys.* **28**, 595–656 (1979).

8. Bogdanov, A. & Hubert, A. Thermodynamically stable magnetic vortex states in magnetic crystals. *J. Magn. Magn. Mater.* **138**, 255–269 (1994).

9. Je, S.-G. *et al.* Direct Demonstration of Topological Stability of Magnetic Skyrmions via Topology Manipulation. *ACS Nano* **14**, 3251–3258 (2020).

10. Sampaio, J., Cros, V., Rohart, S., Thiaville, A. & Fert, A. Nucleation, stability and current-induced motion of isolated magnetic skyrmions in nanostructures. *Nat. Nanotechnol.* **8**, 839–844 (2013).

11. Jiang, W. *et al.* Blowing magnetic skyrmion bubbles. *Science* **349**, 283–286 (2015).

12. Woo, S. *et al.* Observation of room-temperature magnetic skyrmions and their current-driven dynamics in ultrathin metallic ferromagnets. *Nat. Mater.* **15**, 501–506 (2016).

13. Lemesh, I. *et al.* Current-Induced Skyrmion Generation through Morphological Thermal Transitions in Chiral Ferromagnetic Heterostructures. *Adv. Mater.* **30**, 1805461 (2018).

14. Dohi, T., Reeve, Robert. M. & Kläui, M. Thin Film Skyrmionics. *Annu. Rev. Condens. Matter Phys.* **13**, 73-95 (2022).

15. Zázvorka, J. *et al.* Thermal skyrmion diffusion used in a reshuffler device. *Nat. Nanotechnol.* **14**, 658–661 (2019).

16. Raab, K. *et al.* Brownian reservoir computing realized using geometrically confined skyrmion dynamics. *Nat. Commun.* **13**, 6982 (2022).

17. Prychynenko, D. *et al.* Magnetic Skyrmion as a Nonlinear Resistive Element: A Potential Building Block for Reservoir Computing. *Phys. Rev. Appl.* **9**, 014034 (2018).

18. Pinna, D. *et al.* Skyrmion Gas Manipulation for Probabilistic Computing. *Phys. Rev. Appl.* **9**, 064018 (2018).

19. Jibiki, Y. *et al.* Skyrmion Brownian circuit implemented in continuous ferromagnetic thin film. *Appl. Phys. Lett.* **117**, 082402 (2020).

20. Song, K. M. *et al.* Skyrmion-based artificial synapses for neuromorphic computing. *Nat. Electron.* **3**, 148–155 (2020).

21. Brems, M. A., Kläui, M. & Virnau, P. Circuits and excitations to enable Brownian token-based computing with skyrmions. *Appl. Phys. Lett.* **119**, 132405 (2021).

22. Fert, A., Cros, V. & Sampaio, J. Skyrmions on the track. *Nat. Nanotechnol.* **8**, 152–156 (2013).

23. Li, S. *et al.* Magnetic skyrmions for unconventional computing. *Mater. Horiz.* **8**, 854–868 (2021).

24. Grollier, J., Querlioz, D. & Stiles, M. D. Spintronic Nanodevices for Bioinspired Computing. *Proc. IEEE* **104**, 2024–2039 (2016).

25. Göbel, B. & Mertig, I. Skyrmion ratchet propagation: utilizing the skyrmion Hall effect in AC racetrack storage devices. *Sci. Rep.* **11**, 3020 (2021).

26. Yu, X. Z. *et al.* Real-space observation of a two-dimensional skyrmion crystal. *Nature* **465**, 901–904 (2010).

27. Wang, X. S., Yuan, H. Y. & Wang, X. R. A theory on skyrmion size. *Commun. Phys.* **1**, 31 (2018).

28. Miltat, J., Rohart, S. & Thiaville, A. Brownian motion of magnetic domain walls and skyrmions, and their diffusion constants. *Phys. Rev. B* **97**, 214426 (2018).

29. Miki, S. *et al.* Brownian Motion of Magnetic Skyrmions in One- and Two-Dimensional Systems. *J. Phys. Soc. Jpn.* **90**, 083601 (2021).

30. Troncoso, R. E. & Núñez, Á. S. Brownian motion of massive skyrmions in magnetic thin films. *Ann. Phys.* **351**, 850–856 (2014).

31. Zhang, X. *et al.* Skyrmion-skyrmion and skyrmion-edge repulsions in skyrmion-based racetrack memory. *Sci. Rep.* **5**, 7643 (2015).

32. Song, C. *et al.* Commensurability between Element Symmetry and the Number of Skyrmions Governing Skyrmion Diffusion in Confined Geometries. *Adv. Funct. Mater.* **31**, 2010739 (2021).

33. Ge, Y. *et al.* Constructing coarse-grained skyrmion potentials from experimental data with Iterative Boltzmann Inversion. *Commun. Phys.* **6**, 30 (2023).

34. Zázvorka, J. *et al.* Skyrmion Lattice Phases in Thin Film Multilayer. *Adv. Funct. Mater.* **30**, 2004037 (2020).

35. Winkler, T. B. *et al.* Coarse-graining collective skyrmion dynamics in confined geometries. *Appl. Phys. Lett.* **124**, 022403 (2024).

36. Kapfer, S. C. & Krauth, W. Two-Dimensional Melting: From Liquid-Hexatic Coexistence to Continuous Transitions. *Phys. Rev. Lett.* **114**, 035702 (2015).

37. Huang, P. *et al.* Melting of a skyrmion lattice to a skyrmion liquid via a hexatic phase. *Nat. Nanotechnol.* **15**, 761–767 (2020).

38. Reichhardt, C., Ray, D. & Reichhardt, C. J. O. Collective Transport Properties of Driven Skyrmions with Random Disorder. *Phys. Rev. Lett.* **114**, 217202 (2015).

39. Zhang, X. *et al.* Laminar and transiently disordered dynamics of magnetic-skyrmion pipe flow. *Phys. Rev. B* **108**, 144428 (2023).

40. Sutera, S. P. & Skalak, R. The History of Poiseuille's Law. *Annu. Rev. Fluid Mech.* **25**, 1–20 (1993).

41. Thompson, P. A. & Troian, S. M. A general boundary condition for liquid flow at solid surfaces. *Nature* **389**, 360–362 (1997).

42. Vermant, J. & Solomon, M. J. Flow-induced structure in colloidal suspensions. *J. Phys. Condens. Matter* **17**, R187–R216 (2005).

43. Gruber, R. *et al.* Skyrmion pinning energetics in thin film systems. *Nat. Commun.* **13**, 3144 (2022).

44. Isa, L., Besseling, R. & Poon, W. C. K. Shear Zones and Wall Slip in the Capillary Flow of Concentrated Colloidal Suspensions. *Phys. Rev. Lett.* **98**, 198305 (2007).

45. Cloitre, M. & Bonnecaze, R. T. A review on wall slip in high solid dispersions. *Rheol. Acta* **56**, 283–305 (2017).

46. McCord, J. Progress in magnetic domain observation by advanced magneto-optical microscopy. *J. Phys. Appl. Phys.* **48**, 333001 (2015).

47. Litzius, K. *et al.* Skyrmion Hall effect revealed by direct time-resolved X-ray microscopy. *Nat. Phys.* **13**, 170–175 (2017).

48. Litzius, K. *et al.* The role of temperature and drive current in skyrmion dynamics. *Nat. Electron.* **3**, 30–36 (2020).

49. Thiele, A. A. Steady-State Motion of Magnetic Domains. *Phys. Rev. Lett.* **30**, 230–233 (1973).

50. Reichhardt, C., Ray, D. & Reichhardt, C. J. O. Quantized transport for a skyrmion moving on a two-dimensional periodic substrate. *Phys. Rev. B* **91**, 104426 (2015).

51. Brown, B. L., Täuber, U. C. & Pleimling, M. Effect of the Magnus force on skyrmion relaxation dynamics. *Phys. Rev. B* **97**, 020405 (2018).

52. Brown, B. L., Täuber, U. C. & Pleimling, M. Skyrmion relaxation dynamics in the presence of quenched disorder. *Phys. Rev. B* **100**, 024410 (2019).

53. Weißenhofer, M., Rózsa, L. & Nowak, U. Skyrmion Dynamics at Finite Temperatures: Beyond Thiele's Equation. *Phys. Rev. Lett.* **127**, 047203 (2021).

54. Abera Kolech, B. Magnetic Skyrmions and Quasi Particles: A Review on Principles and Applications. in *Vortex Simulation and Identification* (IntechOpen, 2023).

55. Chen, J. C. & Kim, A. S. Brownian Dynamics, Molecular Dynamics, and Monte Carlo modeling of colloidal systems. *Adv. Colloid Interface Sci.* **112**, 159–173 (2004).

56. Unni, H. N. & Yang, C. Brownian dynamics simulation and experimental study of colloidal particle deposition in a microchannel flow. *J. Colloid Interface Sci.* **291**, 28–36 (2005).

57. COMSOL Multiphysics. v. 6.0 www.comsol.com. COMSOL AB, Stockholm, Sweden. (2021).

## Supplementary Materials

**Appendix A. -** The sample stack is specifically designed to exhibit a skyrmion phase slightly above ambient temperature and possesses a low pinning energy landscape, which allows for diffusive skyrmion motion. However, skyrmions can still become pinned which means their positions are more likely to be at certain sites stochastically and their motion is reduced. To pattern the sample, electron beam lithography is used with an electron beam Pioneer system (Raith Nanofabrication) and Argon ion etching with an IonSys Model 500 ion beam etching system. The chromium-gold pads (5 nm Cr and 40 nm Au) are laid out using the lift-off method, and the material is sputtered using a home built sputtering system. Aluminum wires are then wire bonded onto the gold pads of the sample and onto a printed circuit board placed on top of the coil to make the necessary electrical connections. The thicker ends of the patterned wires on the sample serve as reservoirs, providing enough space for skyrmions to nucleate in the required quantity to move through the wire and annihilate in a reservoir in the other end beneath the gold contact. The wires have varying widths, with either straight edges or triangular protrusions at the edges. When applying current densities of $10^{10} \frac{A}{m^2}$ or higher, the skyrmions are pushed with such high velocity that they empty the wire of skyrmions in just a few seconds, and worm domains are pushed in the flow direction into the field of view. Frames with worm domains are manually excluded in the velocity analysis.

**Appendix B. -** To achieve a frame rate of 16 frames per second, the exposure time is set to 62.5 ms and a 2×2 binning technique is used, which averages 4 physical pixels to one virtual pixel in the image. The resulting resolution is 672×512 pixels with a field of view of $(179 \pm 1) \times (136 \pm 1)$ µm. Differential images between the skyrmion and saturation states are employed to enhance the contrast. This results in black and white subtraction errors at the edges of the wire in the images, which could lead to an operational error when the sample/structure moves under the microscope due to thermal drift or mechanical strain. To minimize drift, we used a stable thermal equilibrated state and increased mechanical stiffness. The sample is positioned on a QC-17-1.0-2.5MS Peltier element to achieve temperatures of approximately 315 K at ambient air with a temperature stability of better than 0.3 K. The temperature is measured by a Pt100 resistive heat sensor and allows to realize the skyrmion phase and achieve the appropriate sizes and amounts of skyrmions. In this study, the nucleation of skyrmions is achieved through the application of an out-of-plane (OOP) field in the µT regime and an in-plane field of 25 mT. The number of nucleated skyrmions is dependent on the OOP field, which has an influence on the skyrmion radius. Although the skyrmions exhibit thermal motion, this is largely restricted by both skyrmion-skyrmion repulsion in the dense state and the skyrmion-edge repulsion. The electric current for skyrmion motion is applied using current sources (Keithley 2400 SourceMeter), resulting in a current density of $5.7 \cdot 10^7 \frac{A}{m^2}$ in the valley-to-valley section, with variable wire widths. To analyse the videos, we track the skyrmions in the wire using the Trackmate plugin of ImageJ's Fiji. Some positions during current application and its resulting skyrmion flow are more likely and more persistently filled by a skyrmion, indicating the existence of pinning sites, on which it is energetically favorable for the skyrmion to stay.

**Appendix C. -** In simulations, the random force $\vec{F}_R$ is generated by drawing a uniformly distributed angle φ and a uniformly distributed magnitude $|\vec{F}_R|$ such that the properties of Brownian motion are fulfilled. Namely, $\langle \vec{F}_R \rangle = 0$ and $\langle |\vec{F}_R|^2 \rangle = 2dk_bT \gamma/\delta t$, where d = 2 is the number of dimensions, $k_bT = 1$ is the thermal energy, and $\delta t = 10^{-4}$ is the time step used in simulations. The skyrmion-skyrmion and skyrmion-boundary interactions are purely repulsive and both follow an exponential form: $U(r) = ae^{-r/b}$, where r is the distance between two skyrmions or a skyrmion and the closest point of a boundary, and a and b are based on fit parameters previously determined using iterative Boltzmann inversion[33]. The fit parameters are $a_{sksk} = 735.1 kT$, $b_{SkSk} = 1.079 \mu m$ for the skyrmion-skyrmion interaction, and $a_{SW} = 176.7 kT, b_{SW} = 1.673 \mu m$ for the skyrmion-boundary interaction. The strength of the skyrmion-boundary interaction is halved in simulations with triangular modulation since the potentials were determined for skyrmions in a sample with straight boundaries, while two boundaries can contribute to the skyrmion-boundary interaction within a sample with triangular modulation. Pinning effects are not included in the simulation. Simulations for Fig. 2 were performed using a driving force of $F_D$ = 12.5, whereas backflow in Fig. 3 is observed at $F_D$ = 1.

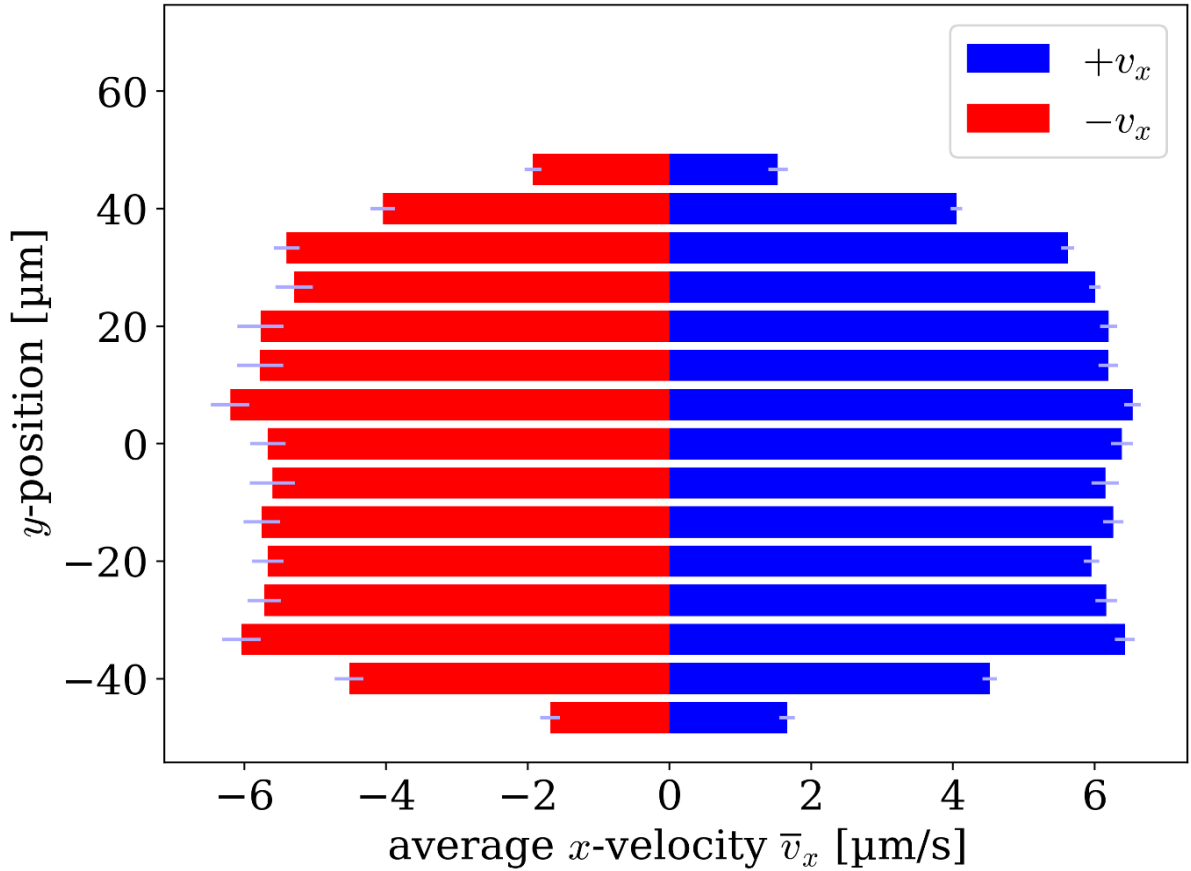

**Supplementary Figure 1:** Velocity profiles of a modulated wire with skyrmion flow in opposite directions. Blue shows the velocity profile for +25 µA applied electric current, while red represents the profile for -25 µA. These currents correspond to a current density of $3.1 \cdot 10^7 \frac{A}{m^2}$ in the middle section. This shows that asymmetries in experimental flow profiles likely originate from the random pinning landscape, since the blue and red profiles can be mapped to each other upon a change of the sign; such a mapping would not be possible if the skyrmion Hall effect was the origin of asymmetries.

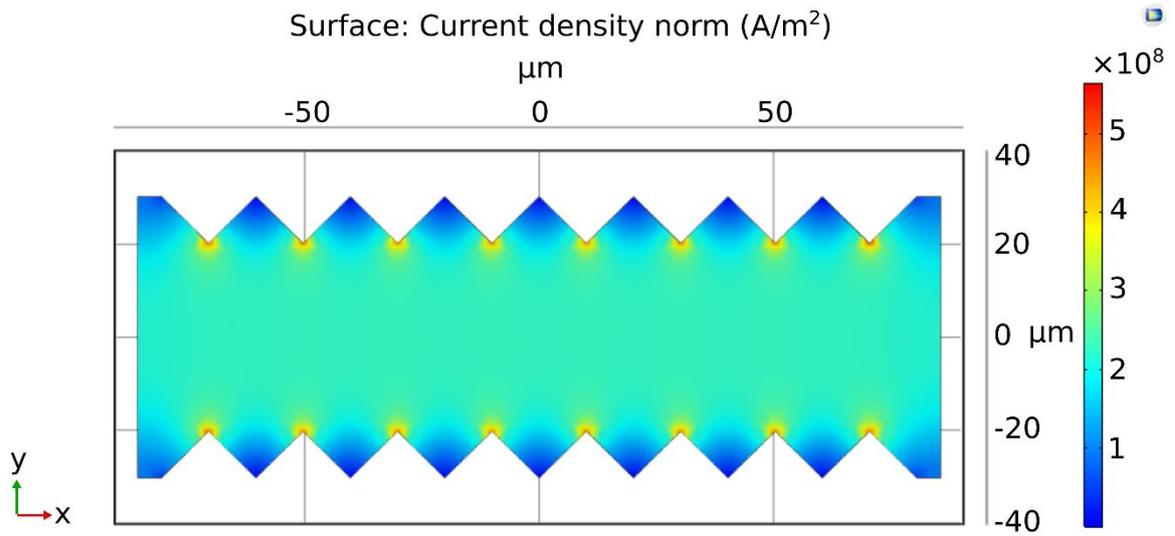

**Supplementary Figure 2:** COMSOL simulation of the current distribution in a 40µm wide modulated channel. The current density in the middle section of the channel can be considered constant, while it becomes highest at the inward-pointing spikes.